\documentclass[12pt]{article}
\usepackage{amsmath,amssymb,amsfonts,amsthm,bbm,mathrsfs,natbib,appendix}
\usepackage{cancel,sectsty,color,enumitem,array,graphicx}
\usepackage[doublespacing]{setspace}
\usepackage[tmargin=1in,bmargin=1in,lmargin=1.25in,rmargin=1.25in]{geometry}
\usepackage[justification=centering]{caption}
\usepackage{hyperref}
\hypersetup{pdfborder = {0 0 0},colorlinks=true,linkcolor=blue,urlcolor=blue,citecolor=blue}

\newtheorem{theorem}{Theorem}

\newtheorem{lemma}{Lemma}
\newtheorem*{assumption_SRD}{Assumption SRD}

\theoremstyle{definition}
\newtheorem{remark_tmp}{Remark}

\theoremstyle{definition}
\newtheorem{example_tmp}{Example}

\allowdisplaybreaks[1]

\newcommand{\forloop}[5][1]{\setcounter{#2}{#3}\ifthenelse{#4}{#5\addtocounter{#2}{#1}\forloop[#1]{#2}{\value{#2}}{#4}{#5}}{}}

\newcommand{\I}{\mathbbm{1}}

\newcommand{\pto}{\to_\mathbb{P}}
\newcommand{\dto}{\to_{\text{d}}}

\newcommand{\E}{\mathbb{E}}
\newcommand{\V}{\mathbb{V}}
\renewcommand{\P}{\mathbb{P}}

\newcommand{\bX}{\mathbf{X}}
\newcommand{\bZ}{\mathbf{Z}}

\newcommand{\bP}{\mathbf{P}}
\newcommand{\bs}{\mathbf{s}}

\newcommand{\bgamma}{\boldsymbol{\gamma}}
\newcommand{\btau}{\boldsymbol{\tau}}

\newcommand{\bmu}{\boldsymbol{\mu}}
\newcommand{\bsigma}{\boldsymbol{\sigma}}
\newcommand{\bSigma}{\boldsymbol{\Sigma}}

\begin{document}

\setcounter{page}{0}

\title{Regression Discontinuity Designs Using Covariates\thanks{We thank the co-Editor, Bryan Graham, and three reviewers for comments. We also thank Stephane Bonhomme, Ivan Canay, David Drukker, Kosuke Imai, Michael Jansson, Lutz Kilian, Pat Kline, Xinwei Ma, Andres Santos, Gonzalo Vazquez-Bare, and participants at the 2017 Annual Conference of the Society for Political Methodology for thoughtful comments and suggestions. Cattaneo gratefully acknowledges financial support from the National Science Foundation through grants SES-1357561 and SES-1459931, and Titiunik gratefully acknowledges financial support from the National Science Foundation through grant SES-1357561. Companion software is available at \url{https://sites.google.com/site/rdpackages/}.
}
}
\author{Sebastian Calonico\thanks{Department of Economics, University of Miami.} \and
Matias D. Cattaneo\thanks{Department of Economics and Department of Statistics, University of Michigan.} \and
Max H. Farrell\thanks{Booth School of Business, University of Chicago.} \and
Roc\'{i}o Titiunik\thanks{Department of Political Science, University of Michigan.}}
\date{May 25, 2018}
\maketitle\thispagestyle{empty}

\newpage
\begin{abstract}
We study regression discontinuity designs when covariates are included in the estimation. We examine local polynomial estimators that include discrete or continuous covariates in an additive separable way, but without imposing any parametric restrictions on the underlying population regression functions. We recommend a covariate-adjustment approach that retains consistency under intuitive conditions, and characterize the potential for estimation and inference improvements. We also present new covariate-adjusted mean squared error expansions and robust bias-corrected inference procedures, with heteroskedasticity-consistent and cluster-robust standard errors. An empirical illustration and an extensive simulation study is presented. All methods are implemented in \texttt{R} and \texttt{Stata} software packages.\bigskip
\end{abstract}

\bigskip


\newpage

\section{Introduction}\label{sec:intro}

The Regression Discontinuity (RD) design is widely used in Economics, Political Science, and many other social, behavioral, biomedical, and statistical sciences. Within the causal inference framework, the RD design is considered to be one of the most credible non-experimental strategies because it relies on weak and easy-to-interpret nonparametric identifying assumptions, which permit flexible and robust estimation and inference for local treatment effects. The key feature of the design is the existence of a score, index, or running variable for each unit in the sample, which determines treatment assignment via hard-thresholding: all units whose score is above a known cutoff are offered treatment, while all units whose score is below this cutoff are not. Identification, estimation, and inference proceed by comparing the responses of units near the cutoff, taking those below (comparison group) as counterfactuals to those above (treatment group). For literature reviews and practical introductions, see \citet*{Imbens-Lemieux_2008_JoE}, \citet*{Lee-Lemieux_2010_JEL}, \citet*{Cattaneo-Escanciano_2017_AIE}, \citet*{Cattaneo-Titiunik-VazquezBare_2017_JPAM}, \citet*{Cattaneo-Idrobo-Titiunik_2018_Book1,Cattaneo-Idrobo-Titiunik_2018_Book2}, and references therein.

Nonparametric identification of the RD treatment effect typically relies on continuity assumptions, which motivate nonparametric local polynomial methods tailored to flexibly approximate, above and below the cutoff, the unknown conditional mean function of the outcome variable given the score. In practice, researchers often choose a local linear polynomial and perform estimation using weighted linear least squares, giving higher weights to observations close to the cutoff. These estimates are then used to assess whether there is a discontinuity in levels, derivatives, or ratios thereof, at the cutoff. If present, this discontinuity is interpreted as some average response to the treatment (assignment) at the cutoff, depending on the setting and assumptions under consideration. 

When practitioners employ weighted least squares to estimate RD effects, they often augment their estimation models with additional predetermined covariates such as demographic or socio-economic characteristics of the units. Despite the pervasiveness of this practice, there has been little formal study of the consequences of covariate adjustment for identification, estimation, and inference of RD effects under standard smoothness conditions and when employing standard nonparametric local polynomial estimators (e.g., local linear regression). This has led to the proliferation of ad-hoc covariate-adjustment practices that, at best, reduce the transparency of the estimation and, at worst, result in generally incomparable, or even inconsistent, estimators. We provide results that formalize and justify covariate adjustment in local polynomial analysis of RD designs, offering valid and practical estimation and inference procedures.

We augment the standard local polynomial regression to allow covariate adjustment in an additive separable, linear-in-parameters way, following very closely the estimation models typically employed by applied researchers. Our approach allows for the inclusion of continuous, discrete, and mixed regressors, and does not require additional smoothing methods such as choosing other bandwidths or kernels, nor parametric functional form assumptions. Importantly, although we consider estimation models where the covariates enter linearly (in parameters), this is purely a local linear projection, and is in no way an assumption about the functional form of the underlying regression functions. In other words, we study the limiting behavior of local linear least squares estimators of regression functions in a fully nonparametric sense. Within this framework, there does not appear to be broad consensus in applied work as to how exactly the covariates should enter the model. We characterize not only the conditions under which the inclusion of covariates is appropriate, but also the ways in which adjusting by covariates may lead to inconsistent or poorly behaved RD estimators. Thus, our formal results offer concrete guidelines for applied researchers who want to employ covariate-adjusted local regression methods, which were previously unavailable.

We advocate the use of a simple covariate-adjusted RD estimator that imposes the same adjustment above and below the cutoff. This estimator is easy to implement, and is consistent for the standard RD treatment effect under the mild, intuitive condition that the treatment has no mean effect on the covariates at the cutoff. For example, in the sharp RD design, the only requirement is that the covariates have equal conditional expectation limits from above and below at the cutoff, which is often conceived and presented as a falsification or ``placebo'' test in RD empirical studies \citep{Lee_2008_JoE}. For kink RD designs, we show that additional conditions on the covariates are needed, suggesting new specification tests for empirical work (all details are provided in the supplemental appendix). The requirement of ``balanced'' covariates at the cutoff is the most natural and practically relevant sufficient condition. We also show that employing other covariate-adjusted estimators may lead to inconsistency and/or invalid inference.

We offer a complete large sample analysis of our recommended covariate-adjusted RD estimator, including novel mean squared error (MSE) expansions, MSE-optimal bandwidths (with consistent data-driven implementations), MSE-optimal point estimators, and valid asymptotic inference, covering all empirically relevant RD designs (sharp RD, kink RD, fuzzy RD, and fuzzy kink RD), with both heteroskedastic and clustered data. We characterize precisely the potential for efficiency gains, which are guaranteed when the best linear effect of the additional covariates on the outcome, at the cutoff, is equal for both control and treatment groups. These results have immediate practical use in any RD analysis and aid in interpreting prior results. We illustrate our methods numerically by revisiting the data of \citet*{Ludwig-Miller_2007_QJE} to reanalyze the effect of Head Start on child mortality, and with an extensive simulation study. We also provide \texttt{Stata} and \texttt{R} packages that implement all our methods \citep*{Calonico-Cattaneo-Farrell-Titiunik_2017_Stata}.

Our framework for covariate adjustment is best suited for settings where the inclusion of covariates is intended to increase the precision of the RD treatment effect estimator, in the same spirit as (pre-intervention) covariates are often included in the analysis of randomized experiments. We require that the treatment have no effect on the covariates at the cutoff, and the conditional expectations of potential outcomes and covariates given the score be continuous at the cutoff. We impose no restrictions on the ``long'' conditional expectation of potential outcomes given the score and the covariates. These conditions are in perfect agreement with nonparametric identification results in the RD literature and with standard regression adjustment arguments in the context of randomized experiments.

An alternative motivation for the inclusion of covariates in RD analysis is to increase the plausibility of the design in cases where researchers fear that the potential outcomes are discontinuous at the cutoff; in these cases, the inclusion of covariates is intended to restore identification of some RD parameter. This second motivation generally requires either parametric assumptions on the regression functions to enable extrapolation, or other design assumptions that redefine the parameter of interest in a fixed local neighborhood rather than at the cutoff point. In this context, our estimation and inference results can continue to be used under stronger assumptions. For example, if the covariate specification in the ``long'' regression functions is assumed correct within a neighborhood around the cutoff, covariate-adjustment leads to identification, estimation, and inference of the corresponding RD treatment effect, as it occurs in standard parametric linear regression settings. However, our main goal here is on nonparametric identification, estimation, and inference of the canonical RD parameter at the cutoff, which is assumed to be well defined to begin with. We plan to investigate covariate-adjustment for identification of other RD treatment effects in future work.

Our paper contributes to the large and still rapidly expanding methodological literature on RD designs---we stay away from summarizing this literature due to space constrains, and refer the reader to the references cited in our opening paragraph. From the specific perspective of covariate adjustment, our paper is related to two strands of the literature. First, a portion of the RD literature re-interprets the data as being ``as good as randomized'' within a small window around the cutoff, though this requires stronger conditions than just continuity/smoothness of conditional expectations as we maintain here \citep*{Lee_2008_JoE,Cattaneo-Frandsen-Titiunik_2015_JCI,Cattaneo-Titiunik-VazquezBare_2017_JPAM,Sekhon-Titiunik_2017_AIE}. Adopting a local randomization perspective, our work is related in obvious ways to the large literature on covariate adjustment in randomized experiments \citep[e.g.,][and references therein]{Imbens-Rubin_2015_Book}, and shows two interesting connections. On the one hand, we find that the conditions under which adjusting for pre-treatment covariates can lead to efficiency gains parallel those required in classical experiments (Section \ref{sec:asympeff}), although we also find that some approaches have inferior estimation and inference properties because of the intrinsic smoothing in RD methods (Lemma \ref{lem:srd-cov}), which is absent in experimental methods. On the other hand, it follows from our results that adjusting for imbalanced covariates in order to restore identification is not possible without functional form assumptions, just like in randomized experiments it is not generally possible to remove imbalances by simply adjusting for the covariates that are correlated with treatment.

Second, \citet*{Frolich-Huber_2018_JBES} develop a nonparametric kernel method to account for observed covariates for RD treatment effect estimation. This localize-then-smooth method of adding covariates to RD designs involves nonparametric estimation over both the running variable and the covariates, and thus requires $d+2$ bandwidth choices (where $d$ denotes the dimension of the additional covariates included). Their estimator recovers a weighted average of local treatment effects, which coincides with the standard RD treatment effect of interest under further ``balance'' type conditions on the covariates. In contrast, our methods do not require any smoothness assumptions on, or nonparametric smoothing over, the ``long'' conditional expectation depending on the $d$ additional covariates, and are in perfect agreement with empirical practice (i.e., local regression methods adding covariates linearly and employing only linear projection arguments). Furthermore, we provide optimal bandwidth selectors and robust bias-corrected inference with valid variance estimators under both heteroskedasticity and clustering, and also cover kink RD designs in addition to sharp and fuzzy designs.

The upcoming presentation is organized as follows. Section \ref{sec:setup} introduces the framework, Sections \ref{sec:id} and \ref{sec:theory} summarize our main methodological results, Section \ref{sec:numerical} briefly discusses numerical results, and Section \ref{sec:conclusion} concludes. Due to space limitations, we discuss only local linear estimation in sharp RD designs, but in the lengthy supplemental appendix we extend our results to fuzzy and kink RD designs for any local polynomial order, in addition to reporting several other results of interest. Companion software, replication files, and other materials can be found at \url{https://sites.google.com/site/rdpackages/}.

\section{Sharp RD Designs Using Covariates}\label{sec:setup}

The observed data is assumed to be a random sample $(Y_i,T_i,X_i,\bZ_i')'$, $i=1,2,\dots,n$. The key feature of any RD design is the presence of a continuous score or running variable $X_i$, with known threshold $\bar{x}$, which determines treatment assignment for each unit in the sample. For simplicity, here we discuss only sharp RD designs, where all units with $X_i \geq \bar{x}$ are assigned to the treatment group and receive the treatment, and all units with $X_i<\bar{x}$ are assigned to the control group and do not receive the treatment. We also normalize $\bar{x}=0$ to save notation. All other cases are discussed in the supplemental appendix. 

Using the potential outcomes framework, the observed outcome is $Y_i=Y_i(0)\cdot(1-T_i)+Y_i(1)\cdot T_i$, where $T_i=\I(X_i\geq\bar{x})$ denotes treatment status and $Y_i(1)$ and $Y_i(0)$ are the potential outcomes with and without treatment, respectively, for each unit $i$ in the sample. The parameter of interest is the average treatment effect at the cutoff:
\[\tau = \E[Y_i(1)-Y_i(0)|X_i=0].\]
Evaluation points of functions are dropped whenever possible throughout the paper. \citet*{Hahn-Todd-vanderKlaauw_2001_ECMA} give precise, easy-to-interpret conditions for nonparametric identification of the standard RD treatment effect $\tau$, without additional covariates. The key substantive identifying assumption is that $\E[Y_i(t)|X_i=x]$, $t\in\{0,1\}$, are continuous at the cutoff $x=0$. 

The new feature studied in this paper is the presence of additional covariates, collected in the random vector $\bZ_i\in\mathbb{R}^d$, which could be continuous, discrete, or mixed. We set $\bZ_i=\bZ_i(0)\cdot(1-T_i)+\bZ_i(1)\cdot T_i$, where $\bZ_i(1)$ and $\bZ_i(0)$ denote the potential covariates on either side of the threshold. In practice, it is natural to assume that some features of the marginal distributions of $\bZ_i(1)$ and $\bZ_i(0)$ are equal near the cutoff or, more generally, that $\bZ_i(1)=_d\bZ_i(0)$, which is implied by the definition of a ``pre-treatment'' or ``pre-determined'' covariate in the causal inference literature.

A large portion of the literature on estimation and inference in RD designs focuses on local polynomial estimators. In practice, researchers first choose a neighborhood around the cutoff, usually via a bandwidth choice $h$, and then conduct local linear polynomial inference---that is, they rely on linear regression fits using only units whose scores $X_i$ lay within that pre-selected neighborhood, with the weighting scheme determined by the choice of kernel function $K(\cdot)$. The role of the kernel and bandwidth is to localize the regression fit near the cutoff. The most popular choices are the uniform kernel (equally weighting observations $X_i\in[-h,h]$) and the triangular kernel (linearly down-weighting observations $X_i\in[-h,h]$). To be specific, the standard local linear RD treatment effect estimator $\hat{\tau}$ is obtained by running the weighted least squares regression of $Y_i$ on a constant, $T_i$, $X_i$, and $X_iT_i$ using only units with $X_i\in[-h,h]$ and applying weights $K(X_i/h)$, which in this paper we denote by
\begin{equation}\label{eqn:RD-standard}
\hat{\tau} \quad:\quad \hat{Y}_i = \hat{\alpha} + T_i\hat{\tau} + X_i\hat{\beta}_- + T_iX_i\hat{\beta}_+.
\end{equation}

We give a precise definition of $\hat{\tau}$ and all other estimators in the supplemental appendix. The estimator $\hat{\tau}$ is, of course, numerically equivalent to the difference in intercepts that would be obtained from two separate weighted least squares regressions using observations on each side of the cutoff (with the same kernel and bandwidth). We set the problem as a single joint least squares linear regression fit to ease the upcoming comparisons with covariate-adjusted RD estimators. We maintain the local linear specification for simplicity but, as shown in the supplemental appendix, all our results apply to any $p$-th order local polynomial fit.

While the standard estimator $\hat{\tau}$ is popular in empirical work, it is common to augment the specification with the additional covariates $\bZ_i$. In practice, these covariates can be included in many different ways. We consider five linear covariate-adjustment specifications, some with centered covariates, to mimic the variety of current empirical practice: 
\begin{align}
\tilde{\tau} \quad:\quad
& \tilde{Y}_i = \tilde{\alpha} + T_i\tilde{\tau} + X_i\tilde{\beta}_- + T_iX_i\tilde{\beta}_+ + \bZ_i'\tilde{\bgamma},
\label{eqn:RD-Cov-commoncoeff}\\
\check{\tau} \quad:\quad
& \check{Y}_i = \check{\alpha} + T_i\check{\tau} + X_i\check{\beta}_- + T_iX_i\check{\beta}_+ + (1-T_i)\bZ_i'\check{\bgamma}_- + T_i\bZ_i'\check{\bgamma}_+,
\label{eqn:RD-Cov-twocoeff}\\
\dot{\tau} \quad:\quad
& \dot{Y}_i = \dot{\alpha} + T_i\dot{\tau} + X_i\dot{\beta}_- + T_iX_i\dot{\beta}_+ + (\bZ_i-\bar{\bZ})'\dot{\bgamma},
\label{eqn:RD-Cov-commoncoeffdemeaned}\\
\ddot{\tau} \quad:\quad
& \ddot{Y}_i = \ddot{\alpha} + T_i\ddot{\tau} + X_i\ddot{\beta}_- + T_iX_i\ddot{\beta}_+ + (1-T_i)(\bZ_i-\bar\bZ)'\ddot{\bgamma}_- + T_i(\bZ_i-\bar\bZ)'\ddot{\bgamma}_+,
\label{eqn:RD-Cov-twocoeffcommondemeaned}\\
\overset{\dots}{\tau} \quad:\quad
& \overset{\dots}{Y}_i = \overset{\dots}{\alpha} + T_i\overset{\dots}{\tau} + X_i\overset{\dots}{\beta}_- + T_iX_i\overset{\dots}{\beta}_+ + (1-T_i)(\bZ_i-\bar\bZ_{-})'\overset{\dots}{\bgamma}_- + T_i(\bZ_i-\bar\bZ_{+})'\overset{\dots}{\bgamma}_+,
\label{eqn:RD-Cov-twocoeffdemeaned}
\end{align}
where all these weighted least squares regressions are computed only for observations with $X_i\in[-h,h]$ and weights $K(X_i/h)$, and $\bar\bZ$, $\bar\bZ_{-}$ and $\bar\bZ_{+}$ correspond, respectively, to the sample average of $Z_i$ for $X_i\in[-h,h]$, $X_i\in[-h,0)$ and $X_i\in[0,h]$.

Our recommended approach is $\tilde{\tau}$ in \eqref{eqn:RD-Cov-commoncoeff}, which we call the \emph{covariate-adjusted RD estimator}. This  estimator broadly captures the common empirical practice of first choosing a neighborhood around the cutoff, and then conducting local linear least squares estimation and inference with covariates. We formalize two important points regarding the way that the additional covariates $\bZ_i$ are used in the local least squares fit: (i) additive separability between the running variable and the covariates, and (ii) linear-in-parameters specification for the covariates. We avoid full nonparametric estimation over $(X_i,\bZ_i')' \in \mathbb{R}^{1+d}$, which would introduce $d$ additional bandwidths and kernels, quickly leading to a curse of dimensionality and hence rendering empirical application infeasible. Furthermore, in practice, $\bZ_i$ could include power expansions, interactions, and other ``flexible'' transformations of the original covariates. This approach to RD covariate adjustment allows for any type of additional regressors, including fixed effects or other discrete variables.

As discussed above, we focus on the common motivation for covariate adjustment based on improving the precision of the estimator of the RD treatment effect, $\tau$, analogously to the standard justification for covariate adjustment in randomized experiments. We build on this intuition and make precise the conditions required for consistency of the covariate-adjusted RD estimator $\tilde{\tau}$ for $\tau$. We also show that much more stringent conditions are required if the estimation model includes treatment-covariate interactions and/or centering.

\subsection{Notation and Regularity Conditions}

The following assumption defines notation and collects the conditions required for sharp RD designs.
\begin{assumption_SRD}
	\phantomsection
	\label{ass:srd}
For $t\in\{0,1\}$ and all $x\in[x_l,x_u]$, where $x_l,x_u\in\mathbb{R}$ such that $x_l<\bar{x}<x_u$:\newline
(a) The Lebesgue density of $X_i$, denoted $f(x)$, is continuous and bounded away from zero.\newline
(b) $\mu_{Y-}(x) = \E[Y_i(0)|X_i=x]$ and $\mu_{Y+}(x) = \E[Y_i(1)|X_i=x]$ are thrice continuously differentiable.\newline
(c) $\bmu_{Z-}(x) = \E[\bZ_i(0)|X_i=x]$ and $\bmu_{Z+}(x) = \E[\bZ_i(1)|X_i=x]$ are thrice continuously differentiable, and $\E[\bZ_i(t)Y_i(t)|X_i=x]$ is continuously differentiable.\newline
(d) $\V[(Y_i(t),\bZ_i(t)')|X_i=x]$ is continuously differentiable and invertible.\newline
(e) $\E[|(Y_i(t),\bZ_i(t)')|^4|X_i=x]$, is continuous, where $|\cdot|$ denotes the Euclidean norm.
\end{assumption_SRD}

Assumption \hyperref[ass:srd]{SRD} imposes standard continuity/smoothness assumptions common to all nonparametric analyzes of RD designs (parts (a) and (b)), plus a mild assumption to allow for the inclusion of additional covariates (part (c)). Parts (d) and (e) are standard restrictions on the conditional variances and higher-moments. Indeed, if one simply ignores all statements involving the additional covariates, the conditions are exactly those found in the RD literature for local linear regression (the supplemental appendix shows the version for polynomial fits of degree $p \geq 1$). 

The assumptions are placed only on features such as the mean and variance of the conditional distributions given the running variable $X_i$ alone. Importantly, Assumption \hyperref[ass:srd]{SRD} does not restrict in any way the ``long'' conditional expectations $\E[Y_i(t)|X_i,\bZ_i(t)]$, $t\in\{0,1\}$, which implies that our methods allow for discrete, continuous, and mixed additional covariates, and do not require any semiparametric or parametric modeling of this regression function. That is, we allow for complete misspecification of $\E[Y_i(t)|X_i,\bZ_i(t)]$ in any finite sample, and hence give a ``best linear approximation'' or ``local linear projection'' interpretation to the RD estimators \eqref{eqn:RD-Cov-commoncoeff}--\eqref{eqn:RD-Cov-twocoeffdemeaned}. In other words, the linearity in these specifications represents the empirical use of the covariates, and does not impose parametric assumptions on the underlying regression functions.

\section{Estimation in Sharp RD Designs Using Covariates}\label{sec:id}
    
We now present the first main result of the paper, which constructs and gives an interpretation to the implicit estimands associated with the covariate-adjusted RD estimators. All limits are taken as $n\to\infty$, unless otherwise noted.

\begin{lemma}[Sharp RD with Covariates]\label{lem:srd-cov}
Let Assumption \hyperref[ass:srd]{SRD} hold, assume the weights obey $K(u)=\I(u<0)k(-u)+\I(u\geq 0)k(u)$, with $k(\cdot):[0,1]\mapsto\mathbb{R}_+$ bounded, zero outside its support, and positive and continuous on $(0,1)$. If $nh\to\infty$ and $h\to0$, then
\begin{align*}
\tilde{\tau} \pto &\; \tau - \left[\bmu_{Z+} - \bmu_{Z-}\right]'\bgamma_Y,\\
\check{\tau} \pto &\; \tau - \left[\bmu_{Z+}'\bgamma_{Y+} - \bmu_{Z-}'\bgamma_{Y-}\right],\\
\dot{\tau}   \pto &\; \tau - \left[\bmu_{Z+} - \bmu_{Z-}\right]'\bgamma_Y,\\
\ddot{\tau}  \pto &\; \tau - \left[(\bmu_{Z+}-\bar\bmu_{Z})'\bgamma_{Y+} - (\bmu_{Z-}-\bar\bmu_{Z})'\bgamma_{Y-}\right],\\
\overset{\dots}{\tau} \pto &\; \tau,
\end{align*}
where $\bgamma_Y = (\bsigma^{2}_{Z-} + \bsigma^{2}_{Z+})^{-1} \E\left[\left.(\bZ_i(0)-\bmu_{Z-}(X_i)) \;Y_i(0) + (\bZ_i(1)-\bmu_{Z+}(X_i)) \;Y_i(1)\right|X_i=\bar{x}\right]$, $\bmu_{Z-} = \bmu_{Z-}(\bar{x})$, $\bgamma_{Y-} = (\bsigma^{2}_{Z-})^{-1} \E\left[\left.(\bZ_i(0)-\bmu_{Z-}(X_i)) \;Y_i(0) \right|X_i=\bar{x}\right]$, $\bsigma^{2}_{Z-} =\V[\bZ_i(0)|X_i=\bar{x}]$, and similarly for $\bmu_{Z+}$, $\bgamma_{Y+}$ and $\bsigma^{2}_{Z+}$, and $\bar\bmu_{Z}=\bmu_{Z+}/2+\bmu_{Z-}/2$.
\end{lemma}

It is well known that under the same conditions imposed here, $\hat{\tau} \pto \tau$. The conclusion of this lemma gives a precise description of the probability limit of each covariate-adjusted sharp RD estimator in \eqref{eqn:RD-Cov-commoncoeff}--\eqref{eqn:RD-Cov-twocoeffdemeaned}. Similar results for other RD designs are shown in the supplemental appendix.

Lemma \ref{lem:srd-cov} shows that our recommended covariate-adjusted RD estimator, $\tilde{\tau}$, is consistent for the standard RD treatment effect at the cutoff, $\tau=\mu_{Y+}-\mu_{Y-}$, plus an additional term that depends on the RD treatment effect on the covariates, $\btau_Z:=\bmu_{Z+}-\bmu_{Z-}$. It follows that a sufficient condition for $\tilde{\tau} \pto \tau$ is that $\bmu_{Z+} = \bmu_{Z-}$, i.e.\ that there is no RD treatment effect on the covariates. This is weaker than assuming that the marginal distributions of $\bZ_i(0)$ and $\bZ_i(1)$ are equal at the cutoff, which is the usual definition of predetermined covariates in randomized experiments.

Further, Lemma \ref{lem:srd-cov} shows that the \textit{treatment-interacted covariate-adjusted RD estimator} $\check{\tau}$, which corresponds to fitting two separate weighted linear regressions on each side of the cutoff, is consistent for $\tau$ under stronger conditions than $\tilde{\tau}$. The difference arises because including this interaction allows $\bgamma_-\neq\bgamma_+$ in the estimation, whereas our recommended covariate-adjusted RD estimator, $\tilde{\tau}$, forces equality. The additional term, $[\bmu_{Z+}'\bgamma_{Y+} - \bmu_{Z-}'\bgamma_{Y-}]$, can be interpreted as the difference of the best linear approximations at the cutoff of the unknown conditional expectations $\E[Y_i(t)|X_i,\bZ_i(0)]$, $t\in\{0,1\}$, based on the additional covariates included in the RD estimation. It follows that a necessary and sufficient condition for $\check{\tau} \pto \tau$ is that $\bmu_{Z+}'\bgamma_{Y+} = \bmu_{Z-}'\bgamma_{Y-}$. However, this is harder to justify in practice than the condition required for the model without the interaction, since we must assume $\bgamma_{Y+} = \bgamma_{Y-}$ in addition to $\bmu_{Z+} = \bmu_{Z-}$ (``covariate balance'') above. From a linear least squares regression perspective, the discrepancy between the probability limits of $\tilde{\tau}$ and $\check{\tau}$ can be explained as a misinterpretation of the interaction term in the estimation model.

Finally, the last three results in Lemma \ref{lem:srd-cov} study different demeaning approaches motivated by the literature on covariate-adjustment in randomized experiments and by classical least squares regression with interactions. As a result of the demeaning of covariates, the probability limit of the estimators change: $\dot{\tau}$ (common demeaning) behaves like $\tilde{\tau}$, $\ddot{\tau}$ (common demeaning, treatment-interaction) has a new probability limit involving a recentering, and $\overset{\dots}{\tau}$ (group-demeaning, treatment-interaction) behaves like $\hat{\tau}$. For the former two, the same condition, $\bmu_{Z+} = \bmu_{Z-}$, is required for consistency; $\overset{\dots}{\tau}$ does not require this condition because the demeaned covariates are essentially orthogonal to treatment status. These requirements and conclusions are conceptually similar to zero-correlation assumptions in standard least squares algebra. Despite these similarities, the nonparametric nature of the problem introduces important differences with standard least squares: formal derivations require nonparametric large-sample approximations (otherwise, misspecification biases appear for fixed $n$), and inference results do not follow from standard parametric least squares arguments.

A very important drawback of all three demeaning-based estimators $(\dot{\tau},\ddot{\tau},\overset{\dots}{\tau})$ is that they employ some form of local average of the covariates $(\bar\bZ,\bar\bZ_{-},\bar\bZ_{+})$, which is equivalent to a kernel regression estimator with a uniform kernel. Therefore, inference using the resulting demeaned estimators is severely affected: these estimators exhibit slower convergence rates, new misspecification biases, and additional asymptotic variability when compared to $\tilde{\tau}$ (or even $\check{\tau}$). Thus, in these cases, the analogy with classical least squares regression breaks down. Furthermore, as shown in the supplemental appendix, simple demeaning does not work for other RD designs.

Putting the above together, and from a practical perspective, not only does Lemma \ref{lem:srd-cov} give general, precise, and intuitive characterizations of the probability limits of the various covariate-adjusted RD estimators, but it also has interesting implications for the analysis and interpretation of RD designs using covariates. Most notably, the lemma shows the conditions under which a covariate-adjusted RD estimator is consistent for the standard (causal) RD treatment effect of interest, $\tau$, and by implication, establishes when estimators with and without covariate adjustment are valid (for estimating $\tau$) and can be compared. We conclude that for estimating $\tau$, $\tilde{\tau}$ requires the weakest assumption, $\bmu_{Z+} = \bmu_{Z-}$, while at the same time is not affected by the poor behavior of demeaning. We therefore advocate this estimator for use in applications, and the rest of the paper gives a thorough analysis of its large sample properties and empirical performance.

In the supplemental appendix we extend the results to other RD designs, where we show that a new condition for ``covariate balance'' emerges for kink RD designs---the additional requirement is that $\bmu^{(1)}_{Z+} = \bmu^{(1)}_{Z-}$, where $\bmu^{(1)}_{Z+}$ and $\bmu^{(1)}_{Z-}$ denote the first derivative of the conditional expectations of the covariates at the cutoff for treatment and control, respectively.

\section{Inference in Sharp RD Designs using Covariates}\label{sec:theory}

Estimation and inference in RD designs using local polynomial methods without covariates (i.e. using only $Y_i$ and $X_i$) has been studied in great detail in recent years (references are given in the introduction). In this section, we study the large sample properties of the covariate-adjusted RD estimator $\tilde{\tau}(h)=\tilde{\tau}$, now making the dependence on $h$ explicit, and assuming throughout that $\bmu_{Z+} = \bmu_{Z-}$ in order to maintain the same standard RD treatment effect of interest (Lemma \ref{lem:srd-cov}). We present new MSE expansions, several data-driven optimal bandwidth selectors, valid distributional approximations based on bias-correction techniques, and consistent standard errors for $\tilde{\tau}(h)$. Analogous results for other RD designs and clustered data are given in the supplemental appendix.

We rely on the following representation (valid for each $n$): $\tilde{\tau}(h) = \hat{\tau}(h) - \hat{\btau}_{Z}(h)'\tilde{\bgamma}_{Y}(h)$, where $\hat{\tau}(h)=\hat{\tau}$ and $\tilde{\bgamma}_{Y}(h)=\tilde{\bgamma}_{Y}$ are given in \eqref{eqn:RD-standard} and \eqref{eqn:RD-Cov-commoncoeff}, respectively, and $\hat{\btau}_{Z}(h)$ is a $d$-dimensional vector containing the standard RD treatment effect estimators for each covariate. In other words, each element of $\hat{\btau}_Z(h)$ is constructed using the corresponding covariate as outcome variable in \eqref{eqn:RD-standard}. Using this partial-out representation, it follows that
\[\tilde{\tau}(h) - \tau
  = \bs(h)'\left[\begin{array}{c}\hat{\tau}(h) - \tau\\\hat{\btau}_Z(h)\end{array}\right]
  = \bs'\left[\begin{array}{c}\hat{\tau}(h) - \tau\\\hat{\btau}_Z(h)\end{array}\right]\{1+o_\P(1)\}\]
where $\bs(h)=(1,\tilde{\bgamma}_{Y}(h)')'$ and $\bs=(1,\bgamma_{Y}')'$, and because $\bs(h)\pto\bs$ using the results underlying Lemma \ref{lem:srd-cov} (and, later, we will also use $\bmu_{Z+} = \bmu_{Z-}$, so that $\btau_Z=\mathbf{0}$). The asymptotic analysis proceeds by studying the (joint) large-sample properties of the vector $\hat{\btau}_S(h) := (\hat{\tau}(h),\hat{\btau}_Z(h)')'$ and then taking the linear combination $\bs(h)$ or $\bs$, as appropriate. Note that $\hat{\btau}_S(h)\pto\btau_S:=(\tau,\btau_Z')'$ under the conditions in Lemma \ref{lem:srd-cov}. We give exact details in the supplemental appendix.

\subsection{MSE Expansion and Data-driven Bandwidth Selection}\label{sec:mse}

We first establish a valid asymptotic MSE-type expansion for the covariate-adjusted RD estimator, based on the representation above, which is useful to develop optimal bandwidth choices and optimal point estimators. Further, the bias expressions will be used to develop inference procedures based on robust bias-correction. The object we study is defined as $\mathsf{MSE}[\tilde{\tau}(h)] = \E[(\bs'\hat{\btau}_S(h)-\bs'\btau_S)^2|\bX] = (\mathsf{Bias}[\tilde{\tau}(h)])^2 + \mathsf{Var}[\tilde{\tau}(h)]$, where $\bX=[X_1,X_2,\cdots,X_n]'$, $\mathsf{Bias}[\tilde{\tau}(h)] := \E[\bs'\hat{\btau}_S(h)-\bs'\btau_S|\bX]$, and $\mathsf{Var}[\tilde{\tau}(h)] := \V[\bs'\hat{\btau}_S(h)|\bX]$.

\begin{theorem}[MSE Expansion]\label{thm:mse}
Let the conditions of Lemma \ref{lem:srd-cov} hold. Then
\[\mathsf{MSE}[\tilde{\tau}(h)]
= h^{4} \mathcal{B}_{\tilde{\tau}}(h)^2 \;\{1+o_{\P}(1)\} + \frac{1}{nh} \mathcal{V}_{\tilde{\tau}}(h),
\vspace{-0.1in}
\]
where the precise expressions for all bias and variance terms are given in the supplemental appendix.
\end{theorem}

The bias and variance expressions in Theorem \ref{thm:mse} are different from those available in the literature \citep{Imbens-Kalyanaraman_2012_REStud,Calonico-Cattaneo-Titiunik_2014_ECMA,Arai-Ichimura_2018_QE} due to the presence of the covariates $\bZ_i$. As a consequence, MSE-optimal bandwidth selection and point estimators are different when covariate-adjustment is employed. Bias-correction techniques and standard error constructions are also different, as discussed below.

The leading bias and variance formulas in Theorem \ref{thm:mse} are derived in pre-asymptotic form. For the bias, the random term $\mathcal{B}_{\tilde{\tau}}(h)$ gives a pre-asymptotic stochastic approximation to the conditional bias of the linearized estimator (hence the presence of the $o_{\P}(1)$ term), whereas the variance term $\mathcal{V}_{\tilde{\tau}}(h)$ is simply obtained by a conditional-on-$\bX$ calculation for the linearized estimator. \citet*{Calonico-Cattaneo-Farrell_2018_JASA} prove, using valid Edgeworth expansions, that employing pre-asymptotic approximations when conducting asymptotic inference in nonparametrics can lead to superior performance. Furthermore, fewer unknown features must be characterized and estimated.

The main constants in Theorem \ref{thm:mse} have a familiar form: the bias and variance are, respectively, $\mathcal{B}_{\tilde{\tau}}(h)=\mathcal{B}_{\tilde{\tau}+}(h)-\mathcal{B}_{\tilde{\tau}-}(h)$ and $\mathcal{V}_{\tilde{\tau}}(h)=\mathcal{V}_{\tilde{\tau}-}(h)+\mathcal{V}_{\tilde{\tau}+}(h)$, where each component stems from estimating the unknown regression function on one side of the cutoff. The bias is entirely due to estimating the unknown functions $\mu_{Y-}(\cdot)$ and $\bmu_{Z-}(\cdot)$ for the control group and $\mu_{Y+}(\cdot)$ and $\bmu_{Z+}(\cdot)$ for the treatment group. When the covariates are not included, these constants reduce exactly to those already available in the literature. In the supplemental appendix, we also give the limiting version of the bias and variance constants; that is, we characterize the fixed, real scalars $\mathcal{B}_{\tilde{\tau}}$ and $\mathcal{V}_{\tilde{\tau}}$ that satisfy $(\mathcal{B}_{\tilde{\tau}}(h),\mathcal{V}_{\tilde{\tau}}(h))\pto(\mathcal{B}_{\tilde{\tau}},\mathcal{V}_{\tilde{\tau}})$.

Assuming that $\mathcal{B}_{\tilde{\tau}}\neq 0$, the MSE-optimal bandwidth choice for the local linear covariate-adjusted RD estimator $\tilde{\tau}(h)$ is:
\[\mathfrak{h}_{\tilde{\tau}} = \left[ \frac{\mathcal{V}_{\tilde{\tau}}/n}{4 \mathcal{B}_{\tilde{\tau}}^2}\right]^{1/5}.\]
This choice can be used to construct a consistent and MSE-optimal covariate-adjusted sharp RD point estimator, $\tilde{\tau}(\mathfrak{h}_{\tilde{\tau}}) \pto \tau$, provided that $\btau_Z=\mathbf{0}$. The convergence rate is not affected by $d = \dim(\bZ_i)$ because no nonparametric smoothing is done on the additional covariates.

To construct feasible MSE-optimal bandwidth choices we proceed in the familiar way. For pilot bandwidths $b \to 0$ and $v \to 0$, we can implement $\tilde{\mathfrak{h}}_{\tilde{\tau}} = \left[\frac{\tilde{\mathcal{V}}_{\tilde{\tau}}(v)/n}{4 \tilde{\mathcal{B}}_{\tilde{\tau}}(b)^2}\right]^{1/5}$, where the exact form of the bias estimator, $\tilde{\mathcal{B}}_{\tilde{\tau}}(b)$, and variance estimator, $\tilde{\mathcal{V}}_{\tilde{\tau}}(v)$, are given in the supplemental appendix (we also show results for a generic degree $p\geq 1$, where the optimal bandwidth decays as $n^{-1/(3+2p)}$). Heuristically, these estimators are formed as plug-in versions of the pre-asymptotic formulas obtained in Theorem \ref{thm:mse}. In the supplemental appendix, we show that these feasible versions of the optimal bandwidths are consistent for their infeasible analogues, i.e., $\tilde{\mathfrak{h}}_{\tilde{\tau}}/\mathfrak{h}_{\tilde{\tau}}\pto 1$.

Finally, in the supplemental appendix, we also discuss other MSE-optimal bandwidth selectors, including (i) separate MSE optimizations on either side of the cutoff, (ii) the MSE for the sum rather than the difference of the one-sided estimators, and (iii) several regularized versions of the plug-in bandwidth selectors. In all cases, the decay rate of these bandwidths matches the MSE-optimal choice, but the exact leading constants differ, and these choices may be more stable in finite samples or more robust to situations where the smoothing bias may be small.

\subsection{Asymptotic Efficiency}\label{sec:asympeff}

In addition to finite-sample efficiency considerations, which are well known from the literature on linear least squares, we can give a precise characterization of the effect of introducing covariates in RD estimation on asymptotic efficiency. Using explicit results proven in the supplemental appendix, we can compare the asymptotic variance of the covariate-adjusted estimator $\tilde{\tau}$, denoted by $\mathcal{V}_{\tilde{\tau}}$, to that of the standard RD estimator $\hat{\tau}$, denoted by $\mathcal{V}_{\hat{\tau}}$. This comparison reduces to studying
\[\frac{ \mathcal{V}_{\tilde{\tau}} }{ \mathcal{V}_{\hat{\tau}} } = \frac{\V[Y_i(0)-\bZ_i(0)'\bgamma_Y|X_i=\bar{x}] + \V[Y_i(1)-\bZ_i(1)'\bgamma_Y|X_i=\bar{x}]}{\V[Y_i(0)|X_i=\bar{x}] + \V[Y_i(1)|X_i=\bar{x}]},\]
where $\bgamma_Y$ is given in Lemma \ref{lem:srd-cov}. In general, a definitive ranking is not available because of the linear combination $\bgamma_Y$, which may not be equal to either $\bgamma_{Y_-}$ or $\bgamma_{Y_+}$ given in Lemma \ref{lem:srd-cov}. However, an interesting special case is $\bgamma_Y=\bgamma_{Y_-}=\bgamma_{Y_+}$, in which case $\bgamma_Y$ reduces to the best linear approximation for each group, and therefore $\V[Y_i(t)-\bZ_i(t)'\bgamma_Y|X_i=\bar{x}] \leq \V[Y_i(t)|X_i=\bar{x}]$. This result implies that whenever the effect of the additional covariates on the potential outcomes near the cutoff (via their local linear projections) is (roughly) the same for both control and treatment units, including the covariates may lead to efficiency gains.

The efficiency results above are based on large sample nonparametric approximations using only local linear projections on the additional covariates. Remarkably, however, these results are in perfect agreement with those in the literature on analysis of experiments obtained using Neyman's repeated sampling \citep{Freedman2008_AoAS,Lin2013_AoAS}, where it is also found that incorporating covariates in randomized controlled trials using linear regression leads to efficiency gains only under particular assumptions---for example, using our notation, when $\E[Y_i(t)|\bZ_i(t),X_i=\bar{x}]=a_t+\bZ_i(t)'\mathbf{b}_t$, $t=0,1$, and $\mathbf{b}_0=\mathbf{b}_1$ (note that this parametric assumption implies $\bgamma_Y=\bgamma_{Y_-}=\bgamma_{Y_+}$).

These results also show that $\tilde{\tau}(\mathfrak{h}_{\tilde{\tau}})$ can be a better point estimator in a MSE sense than its counterpart without covariates, $\hat{\tau}(\mathfrak{h}_{\hat{\tau}})$, where $\mathfrak{h}_{\hat{\tau}}$ denotes the MSE-optimal bandwidth choice for the standard RD estimator $\hat{\tau}$ \citep{Imbens-Kalyanaraman_2012_REStud,Calonico-Cattaneo-Titiunik_2014_ECMA,Arai-Ichimura_2018_QE}. Using the explicit formulas, it is easy to give conditions such that $\mathsf{MSE}[\tilde{\tau}(\mathfrak{h}_{\tilde{\tau}})]<\mathsf{MSE}[\hat{\tau}(\mathfrak{h}_{\hat{\tau}})]$ (both have the same rate of decay), although this is not the main goal of our paper. We still recommend that $\hat{\tau}(\mathfrak{h}_{\hat{\tau}})$ be the benchmark point estimator because it relies on minimal identifying assumptions, and thus researchers can incorporate covariates to increase precision relative to it.

\subsection{Asymptotic Distribution and Valid Inference}

To develop valid asymptotic distributional approximations and inference procedures we employ nonparametric robust bias-correction---see  \citet*{Calonico-Cattaneo-Titiunik_2014_ECMA} and \citet*{Calonico-Cattaneo-Farrell_2018_wp,Calonico-Cattaneo-Farrell_2018_JASA}. Inference based on large-sample distribution theory using MSE-optimal bandwidths will suffer from a first-order bias, leading to invalid hypothesis tests and confidence intervals because of misspecification errors near the cutoff. This local smoothing bias involves the bias term in Theorem \ref{thm:mse}, $\mathcal{B}_{\tilde{\tau}}(h)$, which can be estimated and removed.

The bias term $\mathcal{B}_{\tilde{\tau}}(h)$ is known up to the higher-order derivatives of the unknown regression functions, $\mu_{Y-}(\cdot)$, $\bmu_{Z-}(\cdot)$, $\mu_{Y+}(\cdot)$, and $\bmu_{Z+}(\cdot)$, all capturing the misspecification error introduced by the local polynomial approximation. These objects can be estimated nonparametrically---the complete details are available in the supplemental appendix (we replace $\bs$ by $\bs(h)$ for implementation). For our presentation, we simply take as given the bias estimator $\tilde{\mathcal{B}}_{\tilde{\tau}}(b)$ based on a local quadratic regression (in general, a higher-order polynomial than used to form $\tilde{\tau}(h)$) and a preliminary bandwidth $b \to 0$, possibly different from $h$. Then, the bias-corrected covariate-adjusted sharp RD estimator is
\begin{align}\label{eqn:bc}
\tilde{\tau}^{\mathtt{bc}}(h,b) = \tilde{\tau}(h) - h^{2} \tilde{\mathcal{B}}_{\tilde{\tau}}(b)
\end{align}
An empirically useful choice is $b = h$, which is both allowed by our asymptotic theory and has some optimality properties---see \citet*{Calonico-Cattaneo-Farrell_2018_wp,Calonico-Cattaneo-Farrell_2018_JASA} for theoretical results on robust bias-correction. This bias correction approach captures ``flexible'' regression adjustments to account for misspecification in finite samples \citep[][Remark 7]{Calonico-Cattaneo-Titiunik_2014_ECMA}.

The idea behind the robust bias-corrected distributional approximation is to employ an estimator of the variability of $\tilde{\tau}^{\mathtt{bc}}(h,b)$ for Studentization purposes, rather than an estimator of the variability of $\tilde{\tau}(h)$ only. Thus, the final missing ingredient before we can state our asymptotic Gaussianity result is characterizing the (conditional) variance of the bias-corrected covariate-adjusted RD estimator. Its fixed-$n$ variability is easily characterized by
\[
\mathcal{V}_{\tilde{\tau}}^{\mathtt{bc}}(h,b)
= \left[\bs'\otimes\bP^{\mathtt{bc}}_{-}(h,b)\right]\bSigma_{S-}
  \left[\bs'\otimes\bP^{\mathtt{bc}}_{-}(h,b)\right]'
+ \left[\bs'\otimes\bP^{\mathtt{bc}}_{+}(h,b)\right]\bSigma_{S+}
  \left[\bs'\otimes\bP^{\mathtt{bc}}_{+}(h,b)\right]',
\]
where the $n$-vectors $\bP^{\mathtt{bc}}_{-}(h,b)$ and $\bP^{\mathtt{bc}}_{+}(h,b)$ can be computed directly from the data, and the $n(1+d)\times n(1+d)$ matrices of variances and covariances, $\bSigma_{S-}$ and $\bSigma_{S+}$, are the only unknowns. The supplemental appendix collects details and specific formulas.

The (infeasible) variance formula $\mathcal{V}_{\tilde{\tau}}^{\mathtt{bc}}(h,b)$ differs from that presented in Theorem \ref{thm:mse}, $\mathcal{V}_{\tilde{\tau}}(h)$, because it also accounts for the additional variability injected by the bias estimation, $h^{2} \tilde{\mathcal{B}}_{\tilde{\tau}}(b)$. By virtue of the variance formula being computed both conditionally and pre-asymptotically, up to the linear combination term $\bs$, it involves only one unknown feature, $\bSigma_{S-}$ and $\bSigma_{S+}$, which must be estimated, thereby considerably simplifying implementation.

To operationalize the variance formula, we replace unknown quantities by plug-in estimators thereof, which must account for the specific data structure at hand, such as heteroskedasticity or clustering. In the supplemental appendix, we discuss two different plug-in variance estimators, one based on a nearest-neighbor (NN) approach, and the other based on a plug-in residuals (PR) approach, covering both conditional heteroskedasticity and clustered data. Herein, we let $\tilde{\mathcal{V}}_{\tilde{\tau}}^{\mathtt{bc}}(h,b)$ denote a generic, feasible estimator of $\mathcal{V}_{\tilde{\tau}}^{\mathtt{bc}}(h,b)$.

Putting together all the pieces, we obtain the following distributional approximation result which provides valid local polynomial inference in sharp RD designs using covariates.
\begin{theorem}[Asymptotic Normality]\label{thm:normal}
Let the conditions of Theorem \ref{thm:mse} hold, and assume $\btau_Z=0$. If $nh^7 \to 0$ and $\varlimsup \, (h/b) < \infty$, then
\[
	\tilde{T}_{\tilde{\tau}} = \frac{\tilde{\tau}^{\mathtt{bc}}(h,b) - \tau}
	{\sqrt{ \bigl. (nh)^{-1} \bigr. \mathcal{V}_{\tilde{\tau}}^{\mathtt{bc}}(h,b)}} 
	 \dto \mathcal{N}(0,1) 
	  \qquad\quad \text{ and } \qquad\quad   		  \tilde{\mathcal{V}}_{\tilde{\tau}}^{\mathtt{bc}}(h,b)/\mathcal{V}_{\tilde{\tau}}^{\mathtt{bc}}(h,b)\pto 1.
\]
\end{theorem}
Extensions of this result to all other popular RD designs are available in the supplemental appendix. One of the strengths of Theorem \ref{thm:normal} is that the distributional approximation is valid even when the MSE-optimal bandwidth choice is used, which is not true of standard inference procedures. Once bandwidths are chosen, asymptotically valid inference procedures are easily constructed. For example, an approximately $95\%$ robust bias-corrected covariate-adjusted confidence interval for the RD treatment effect $\tau$ using a common bandwidth $h=b$ is given by
\[
\left[\tilde{\tau}^{\mathtt{bc}}(h,h) - \frac{1.96}{\sqrt{nh}} \cdot \sqrt{\tilde{\mathcal{V}}_{\tilde{\tau}}^{\mathtt{bc}}(h,h)} \; , \;
      \tilde{\tau}^{\mathtt{bc}}(h,h) + \frac{1.96}{\sqrt{nh}} \cdot \sqrt{\tilde{\mathcal{V}}_{\tilde{\tau}}^{\mathtt{bc}}(h,h)} \right]. \]

A particularly attractive alternative to MSE-optimal bandwidth selection is to develop coverage error rate (CER) optimal bandwidth choices. Following the valid Edgeworth expansions by \citet*{Calonico-Cattaneo-Farrell_2018_JASA,Calonico-Cattaneo-Farrell_2018_wp}, we also propose the following plug-in bandwidth selector $\tilde{\mathfrak{h}}_{\mathtt{CER},\tilde{\tau}} = n^{-1/20} \times \tilde{\mathfrak{h}}_{\tilde{\tau}}$. This bandwidth choice minimizes the coverage error rate for confidence intervals based on Theorem \ref{thm:normal} below, and may be preferred for inference purposes (the supplemental appendix gives the rate-scaling for generic degree $p \geq 1$.) See \citet*{Cattaneo-VazquezBare_2016_ObsStud} for an introductory discussion on bandwidth selection for RD designs.

Theorems \ref{thm:mse} and \ref{thm:normal} can also be established under clustered sampling. All derivations and results remain valid, but the variance formulas will depend on the particular form of clustering. In this case, asymptotics are conducted under the standard assumptions: (i) each unit $i$ belongs to exactly one of $G$ clusters, and (ii) $G \to \infty$ and $G h \to \infty$---see \cite{Cameron-Miller_2015_JHR} for a review of cluster-robust inference, and \citet*{Bartalotti-Brummet_2017_AIE} for a discussion in the context of MSE-optimal bandwidth selection for sharp RD designs. This extension is conceptually straightforward but notationally cumbersome, and is deferred to the supplemental appendix. Our companion software in \texttt{R} and \texttt{Stata} also includes cluster-robust options for bandwidth selection, MSE-optimal point estimation, and robust bias-corrected inference.

\section{Numerical Results}\label{sec:numerical}

We briefly summarize the main numerical findings from an empirical illustration and a Monte Carlo study. Additional simulation results can be found in the supplemental appendix.

\subsection{Empirical Illustration: Head Start Data}

To illustrate our methods we first re-analyze the effect of Head Start assistance on child mortality in the U.S., which was first studied by \citet*{Ludwig-Miller_2007_QJE}. The unit of observation is the U.S. county, the treatment is receiving technical assistance to apply for Head Start funds, and the running variable is the county-level poverty index constructed in $1965$ by the federal government based on 1960 census information, with cutoff $\bar{x}=59.1984$. The outcome is the child mortality rate (for children of ages five to nine) due to causes affected by Head Start's health services component. We compare the standard RD estimator to the covariate-adjusted RD estimator employing heteroskedasticity-robust nearest-neighbor variance estimation. There are nine pre-intervention covariates from the 1960 U.S. Census: total population, percentage of black and urban population, and levels and percentages of population in three age groups (children aged $3$ to $5$, children aged $14$ to $17$, and adults older than $25$). Full replication code in both \texttt{R} and \texttt{Stata} is available at \url{https://sites.google.com/site/rdpackages/replication/}.

Table \ref{tab:empapp} presents the main results. The first row reports the local linear point estimate using the corresponding MSE-optimal bandwidth $h$ as described in each column. The next three rows report $95\%$ robust bias-corrected confidence intervals, the percentage length change of the covariate-adjusted confidence interval relative to the unadjusted one, and the p-value associated with the hypothesis of zero RD treatment effect. These three rows appear twice, first when $h$ for the RD point estimator and $b$ for the bias estimator are chosen separately, and then when $b=h$. Finally, the last two rows report, respectively, the estimated bandwidths and the number of observations to the left and to the right of the cutoff with $X_i\in[\bar{x}-h,\bar{x}+h]$.

\begin{table}\renewcommand{\arraystretch}{1.2}
	{\begin{center}
			\caption{Empirical Illustration (Head Start Data)}\label{tab:empapp}
			\resizebox{\textwidth}{!}{
\begin{tabular}{lcccc}
\hline\hline
\multicolumn{1}{l}{\bfseries MSE-optimal bandwidths:}&\multicolumn{2}{c}{\bfseries not using covariates}&\multicolumn{1}{c}{\bfseries }&\multicolumn{1}{c}{\bfseries using covariates}\tabularnewline
\cline{2-3} \cline{5-5}
\multicolumn{1}{l}{}&\multicolumn{1}{c}{Standard}&\multicolumn{1}{c}{Cov-adjusted}&\multicolumn{1}{c}{}&\multicolumn{1}{c}{Cov-adjusted}\tabularnewline
\hline
RD treatment effect&$-2.41$&$-2.51$&&$-2.47$\tabularnewline
Inference with $h/b$ unrestricted&$$&$$&&$$\tabularnewline
$\;\;\;\;$Robust 95\% CI&$[\;-5.46\;,\;-0.10\;]$&$[\;-5.37\;,\;-0.45\;]$&&$[\;-5.21\;,\;-0.37\;]$\tabularnewline
$\;\;\;\;$CI length change (\%)&$$&$-8.25$&&$-9.76$\tabularnewline
$\;\;\;\;$Robust p-value&$0.042$&$0.021$&&$0.024$\tabularnewline
Inference with $h/b = 1$&$$&$$&&$$\tabularnewline
$\;\;\;\;$Robust 95\% CI&$[\;-6.41\;,\;-1.09\;]$&$[\;-6.64\;,\;-1.46\;]$&&$[\;-6.54\;,\;-1.39\;]$\tabularnewline
$\;\;\;\;$CI length change (\%)&$$&$-2.86$&&$-3.23$\tabularnewline
$\;\;\;\;$Robust p-value&$0.006$&$0.002$&&$0.003$\tabularnewline
$h\;|\;b$&$6.81 \;|\; 10.72$&$6.81 \;|\; 10.72$&&$6.98 \;|\; 11.64$\tabularnewline
$n_-\;|\;n_+$&$234 \;|\; 180$&$234 \;|\; 180$&&$240 \;|\; 184$\tabularnewline
\hline
\end{tabular}
}
	\end{center}}
	\vspace{.05in}\footnotesize\textbf{Notes}: (i) All estimates are computed using a triangular kernel and nearest neighbor heteroskedasticity-robust variance estimators. (ii) Columns under ``Standard'' and ``Cov-adjusted'' correspond to, respectively, standard and covariate-adjusted RD estimation and inference methods, given a choice of bandwidths. (iii) Bandwidths used ($h$ and $b$) are data-driven MSE-optimal for either standard RD estimator or covariate-adjusted RD estimator (depending on the group of columns). Specifically, in the first two columns (\textit{not using covariates}) the bandwidths are selected to be MSE-optimal for $\hat{\tau}$ (standard RD estimation), while in the third column (\textit{using covariates}) the bandwidths are selected to be MSE-optimal for $\tilde{\tau}$ (covariate-adjusted RD estimation).
\end{table}

The empirical findings are consistent with our theoretical results: employing covariate-adjusted RD inference leads to precision improvements while the point estimators remain stable. The point estimate ranges from $-2.41$ to $-2.51$, and it is statistically different from zero at $5\%$ level in all cases. As should be expected when the additional covariates are truly predetermined, including covariates does not substantially alter the point estimates  (we also implemented ``placebo tests'' on the additional covariates and found, as expected, no statistical evidence of RD treatment effects). Including covariates leads to sizable efficiency gains: the rows labeled ``CI length change (\%)'' show a nearly 10\% efficiency gain when the bandwidths are unrestricted and optimally chosen using covariates.

\subsection{Simulation Evidence}

We also investigate the finite sample performance of our methods using realistic simulated data. We consider four data generating processes constructed using the data of \citet*{Lee_2008_JoE}, where all parameters were obtained from real data unless explicitly noted otherwise. This simulation model has been used extensively in the literature, which facilitates the comparison across studies. All the models include a predetermined covariate (previous democratic vote share), and they vary in the importance of this covariate: (i) in Model 1, the covariate is irrelevant; (ii) in Model 2, it enters the conditional expectation of the potential outcomes $\E[Y_i(t)|X_i,\bZ_i(t)]$, $t\in\{0,1\}$ according to the real data; (iii) Model 3 takes Model 2 but sets the residual correlation between the outcome and covariate to zero; and (iv) Model 4 takes Model 2 but doubles the residual correlation between the outcome and covariate equations. Models 3 and 4 do not imply $\mathbb{C}ov[Y_i(t),\bZ_i(t)|X_i]=0$, $t\in\{0,1\}$.

The constructions allow $\E[Y_i(t)|X_i,\bZ_i(t)]$ to have different coefficients on each side of the cutoff, while the conditional expectation of the potential covariates $\E[\bZ_i(t)|X_i]$, $t\in\{0,1\}$, are constructed assuming they are continuous at the cutoff (but still with different coefficients on either side). Therefore, our covariate-adjusted RD estimator is ``misspecified'' when viewed as a local weighted least squares fit. To conserve space, all details and results of our Monte Carlo study are given in Part V of the supplemental appendix. All findings are consistent with our large sample theory. We find that covariate-adjusted local polynomial analysis can improve both MSE and interval length, sometimes dramatically. The gains are largest in Model 4, with the amplified residual correlation, and least in Model 3, when that channel is shut down, as the theory predicts. The results for Model 1 show that including an irrelevant covariate hardly changes empirical results and conclusions. Finally, we find that our data-driven bandwidth selectors work reasonably well.

\section{Conclusion}\label{sec:conclusion}

We provided a formal framework for identification, estimation, and inference in RD designs when covariates are included in local polynomial estimation. We augmented the standard local polynomial estimator with covariates entering in an additive-separable, linear-in-parameters way, and showed that the resulting covariate-adjusted RD estimator remains consistent for the standard RD treatment effect if the covariate adjustment is restricted to be equivalent above and below the cutoff. Furthermore, this estimator can achieve substantial efficiency gains relative to the unadjusted RD estimator. Thus, we are able to characterize precisely the potential for point estimation and inference improvements, and in particular, efficiency gains. We also provided new MSE expansions, several optimal bandwidth choices and optimal point estimators, robust nonparametric inference procedures based on bias-correction, and heteroskedasticity-consistent and cluster-robust standard errors. Our results and practical methods cover sharp, fuzzy, and kink RD designs, and we also discuss extensions to clustered data. Finally, we illustrated the practical implications of our results using both an application and simulated data.

\clearpage
\bibliographystyle{jasa}
\bibliography{Calonico-Cattaneo-Farrell-Titiunik_2018_RESTAT--Bibliography}

\begin{thebibliography}{24}
\newcommand{\enquote}[1]{``#1''}
\expandafter\ifx\csname natexlab\endcsname\relax\def\natexlab#1{#1}\fi

\bibitem[Arai and Ichimura(2018)]{Arai-Ichimura_2018_QE}
Arai, Y., and Ichimura, H. (2018), \enquote{Simultaneous Selection of Optimal
  Bandwidths for the Sharp Regression Discontinuity Estimator,}
  \emph{Quantitative Economics\emph{, forthcoming}}, 9, 441--482.
\bibitem[Bartalotti and Brummet(2017)]{Bartalotti-Brummet_2017_AIE}
Bartalotti, O., and Brummet, Q. (2017), \enquote{Regression Discontinuity
  Designs with Clustered Data,} in \emph{Regression Discontinuity Designs:
  Theory and Applications (Advances in Econometrics, volume 38)}, eds. M.~D.
  Cattaneo and J.~C. Escanciano, Emerald Group Publishing, pp.\  383--420.
\bibitem[Calonico {\normalfont et~al.}(2018{\natexlab{a}})Calonico, Cattaneo
  and Farrell]{Calonico-Cattaneo-Farrell_2018_wp}
Calonico, S., Cattaneo, M.~D., and Farrell, M.~H. (2018{\natexlab{a}}),
  \enquote{Coverage Error Optimal Confidence Intervals,} \emph{\emph{Working
  paper, University of Michigan}}.
\bibitem[Calonico {\normalfont et~al.}(2018{\natexlab{b}})Calonico, Cattaneo
  and Farrell]{Calonico-Cattaneo-Farrell_2018_JASA}
\leavevmode\vrule height .65ex depth -.6ex width 3em\  (2018{\natexlab{b}}),
  \enquote{On the Effect of Bias Estimation on Coverage Accuracy in
  Nonparametric Inference,} \emph{Journal of the American Statistical
  Association\emph{, forthcoming}}.
\bibitem[Calonico {\normalfont et~al.}(2017)Calonico, Cattaneo, Farrell and
  Titiunik]{Calonico-Cattaneo-Farrell-Titiunik_2017_Stata}
Calonico, S., Cattaneo, M.~D., Farrell, M.~H., and Titiunik, R. (2017),
  \enquote{\texttt{rdrobust}: Software for Regression Discontinuity Designs,}
  \emph{Stata Journal}, 17, 372--404.
\bibitem[Calonico {\normalfont et~al.}(2014)Calonico, Cattaneo and
  Titiunik]{Calonico-Cattaneo-Titiunik_2014_ECMA}
Calonico, S., Cattaneo, M.~D., and Titiunik, R. (2014), \enquote{Robust
  Nonparametric Confidence Intervals for Regression-Discontinuity Designs,}
  \emph{Econometrica}, 82, 2295--2326.
\bibitem[Cameron and Miller(2015)]{Cameron-Miller_2015_JHR}
Cameron, A.~C., and Miller, D.~L. (2015), \enquote{A Practitioner's Guide to
  Cluster-Robust Inference,} \emph{Journal of Human Resources}, 50, 317--372.
\bibitem[Cattaneo and Escanciano(2017)]{Cattaneo-Escanciano_2017_AIE}
Cattaneo, M.~D., and Escanciano, J.~C. (2017), \emph{Regression Discontinuity
  Designs: Theory and Applications (Advances in Econometrics, volume 38)},
  Emerald Group Publishing.
\bibitem[Cattaneo {\normalfont et~al.}(2015)Cattaneo, Frandsen and
  Titiunik]{Cattaneo-Frandsen-Titiunik_2015_JCI}
Cattaneo, M.~D., Frandsen, B., and Titiunik, R. (2015), \enquote{Randomization
  Inference in the Regression Discontinuity Design: An Application to Party
  Advantages in the U.S. Senate,} \emph{Journal of Causal Inference}, 3, 1--24.
\bibitem[Cattaneo {\normalfont et~al.}(2018{\natexlab{a}})Cattaneo, Idrobo and
  Titiunik]{Cattaneo-Idrobo-Titiunik_2018_Book1}
Cattaneo, M.~D., Idrobo, N., and Titiunik, R. (2018{\natexlab{a}}), \emph{A
  Practical Introduction to Regression Discontinuity Designs: Volume I},
  Cambridge Elements: Quantitative and Computational Methods for Social
  Science, Cambridge University Press\emph{, forthcoming}.
\bibitem[Cattaneo {\normalfont et~al.}(2018{\natexlab{b}})Cattaneo, Idrobo and
  Titiunik]{Cattaneo-Idrobo-Titiunik_2018_Book2}
\leavevmode\vrule height .65ex depth -.6ex width 3em\  (2018{\natexlab{b}}),
  \emph{A Practical Introduction to Regression Discontinuity Designs: Volume
  II}, Cambridge Elements: Quantitative and Computational Methods for Social
  Science, Cambridge University Press\emph{, in preparation}.
\bibitem[Cattaneo {\normalfont et~al.}(2017)Cattaneo, Titiunik and
  Vazquez-Bare]{Cattaneo-Titiunik-VazquezBare_2017_JPAM}
Cattaneo, M.~D., Titiunik, R., and Vazquez-Bare, G. (2017), \enquote{Comparing
  Inference Approaches for RD Designs: A Reexamination of the Effect of Head
  Start on Child Mortality,} \emph{Journal of Policy Analysis and Management},
  36, 643--681.
\bibitem[Cattaneo and Vazquez-Bare(2016)]{Cattaneo-VazquezBare_2016_ObsStud}
Cattaneo, M.~D., and Vazquez-Bare, G. (2016), \enquote{The Choice of
  Neighborhood in Regression Discontinuity Designs,} \emph{Observational
  Studies}, 2, 134--146.
\bibitem[Freedman(2008)]{Freedman2008_AoAS}
Freedman, D.~A. (2008), \enquote{On Regression Adjustments in Experiments with
  Several Treatments,} \emph{Annals of Applied Statistics}, 2, 176--196.
\bibitem[Fr\"olich and Huber(2018)]{Frolich-Huber_2018_JBES}
Fr\"olich, M., and Huber, M. (2018), \enquote{Including Covariates in the
  Regression Discontinuity Design,} \emph{Journal of Business \& Economic
  Statistics\emph{, forthcoming}}.
\bibitem[Hahn {\normalfont et~al.}(2001)Hahn, Todd and van~der
  Klaauw]{Hahn-Todd-vanderKlaauw_2001_ECMA}
Hahn, J., Todd, P., and van~der Klaauw, W. (2001), \enquote{Identification and
  Estimation of Treatment Effects with a Regression-Discontinuity Design,}
  \emph{Econometrica}, 69, 201--209.
\bibitem[Imbens and Lemieux(2008)]{Imbens-Lemieux_2008_JoE}
Imbens, G., and Lemieux, T. (2008), \enquote{Regression Discontinuity Designs:
  A Guide to Practice,} \emph{Journal of Econometrics}, 142, 615--635.
\bibitem[Imbens and Kalyanaraman(2012)]{Imbens-Kalyanaraman_2012_REStud}
Imbens, G.~W., and Kalyanaraman, K. (2012), \enquote{Optimal Bandwidth Choice
  for the Regression Discontinuity Estimator,} \emph{Review of Economic
  Studies}, 79, 933--959.
\bibitem[Imbens and Rubin(2015)]{Imbens-Rubin_2015_Book}
Imbens, G.~W., and Rubin, D.~B. (2015), \emph{Causal Inference in Statistics,
  Social, and Biomedical Sciences}, Cambridge University Press.
\bibitem[Lee(2008)]{Lee_2008_JoE}
Lee, D.~S. (2008), \enquote{Randomized Experiments from Non-random Selection in
  U.S. House Elections,} \emph{Journal of Econometrics}, 142, 675--697.
\bibitem[Lee and Lemieux(2010)]{Lee-Lemieux_2010_JEL}
Lee, D.~S., and Lemieux, T. (2010), \enquote{Regression Discontinuity Designs
  in Economics,} \emph{Journal of Economic Literature}, 48, 281--355.
\bibitem[Lin(2013)]{Lin2013_AoAS}
Lin, W. (2013), \enquote{Agnostic Notes on Regression Adjustments to
  Experimental Data: Reexamining Freedman's Critique,} \emph{Annals of Applied
  Statistics}, 7, 295--318.
\bibitem[Ludwig and Miller(2007)]{Ludwig-Miller_2007_QJE}
Ludwig, J., and Miller, D.~L. (2007), \enquote{Does Head Start Improve
  Children's Life Chances? Evidence from a Regression Discontinuity Design,}
  \emph{Quarterly Journal of Economics}, 122, 159--208.
\bibitem[Sekhon and Titiunik(2017)]{Sekhon-Titiunik_2017_AIE}
Sekhon, J., and Titiunik, R. (2017), \enquote{On Interpreting the Regression
  Discontinuity Design as a Local Experiment,} in \emph{Regression
  Discontinuity Designs: Theory and Applications (Advances in Econometrics,
  volume 38)}, eds. M.~D. Cattaneo and J.~C. Escanciano, Emerald Group
  Publishing, pp.\  1--28.
\end{thebibliography}

\clearpage
\begin{appendices}
\section{Supplemental Files}
\small
\onehalfspacing

\begin{itemize}
\item A supplemental appendix, containing the proofs of the main results, several extensions,
additional methodological and technical results, and further simulation details, not included in
the main paper to conserve space, is available \href{https://sites.google.com/site/rdpackages/rdrobust/Calonico-Cattaneo-Farrell-Titiunik_2018_RESTAT--Supplement.pdf}{here}.

\item Replication files are available \href{https://sites.google.com/site/rdpackages/replication}{here}.

\item All methods are implemented in the {\tt rdrobust} package available \href{https://sites.google.com/site/rdpackages/rdrobust}{here}.

\end{itemize}
\end{appendices}

\end{document}